\documentclass[11pt,fleqn,twoside]{article}

\newif\ifreport
\reporttrue

\usepackage[english,german]{babel}
\usepackage{derireport}

\usepackage{alltt}
\usepackage{verbatim}
\usepackage{graphicx}
\usepackage{url}
\usepackage{subfigure}
\usepackage{listings}
\usepackage{color}
\usepackage{lscape}
\usepackage{xspace}
\usepackage{latexsym}
\usepackage{rotating}
\usepackage{enumitem}
\usepackage{amsfonts}
\newcommand{\tickYes}{\checkmark}
\usepackage{pifont}
\newcommand{\tickNo}{\hspace{1pt}\ding{55}}

\foreignreport

\title{Querying over Federated SPARQL Endpoints---A State of the Art Survey}

\authornames{
Nur Aini Rakhmawati
\and
J\"urgen Umbrich \and Marcel Karnstedt \and Ali Hasnain \and  Michael Hausenblas
}

\innertitle{Querying over Federated SPARQL Endpoints---A State of the Art Survey}

\author{
Nur Aini Rakhmawati\affiliation{
DERI, National University of Ireland, Galway, Ireland.
\mbox{firstname.lastname@deri.org}
}
\and
J\"urgen Umbrich\footnotemark[1] \and Marcel Karnstedt\footnotemark[1] \and Ali Hasnain\footnotemark[1]
\and
Michael Hausenblas\footnotemark[1]
}

\acknowledgement{
This work has been carried out in the Linked Data Research Centre (LiDRC) and has been supported by Science Foundation Ireland under Lion 2 and the European Commission's FP7 Support Action LOD-Around-The-Clock (LATC), project no. 256975, Intelligent Information Management (ICT-2009.4.3).
}

\abstract{
The increasing amount of Linked Data and its inherent distributed nature have attracted significant attention throughout the research community and amongst practitioners to search data, in the past years. Inspired by research results from traditional distributed databases, different approaches for managing federation over SPARQL Endpoints have been introduced. SPARQL is the standardised query language for RDF, the default data model used in Linked Data deployments and SPARQL Endpoints are a popular access mechanism provided by many Linked Open Data (LOD) repositories. In this paper, we initially give an overview of the federation framework infrastructure and then proceed with a comparison of existing SPARQL federation frameworks. Finally, we highlight shortcomings in existing frameworks, which we hope helps spawning new research directions.
\\*[\parskip]
~
\\*[\parskip]
}

\deritr{2013-06-07}
\date{June 2013}

\begin{document}

%
%

\maketitle

\newpage
\pagenumbering{Roman}

{\small

\tableofcontents
}


\newpage
\pagenumbering{arabic}

\renewcommand\lstlistingname{Query}    
\renewcommand{\thetable}{\Roman{table}}

\section{Introduction}
\label{intro}

The Resource Description Framework (RDF)\footnote{\url{http://www.w3.org/RDF/}} was introduced in the last decade and now has become a standard for exchanging data in the Web. At present, huge amount of data has been converted to RDF. The SPARQL Protocol and RDF Query Language (SPARQL)\footnote{\url{http://www.w3.org/TR/rdf-sparql-query/}} was officially introduced in 2008 to retrieve RDF data as easily as SQL\footnote{\url{http://www.iso.org/iso/catalogue_detail.htm?csnumber=45498}} does for relational databases. As Web of data grows with more applications rely on it, the number of SPARQL Endpoints constructing SPARQL queries over Web of Data using HTTP also grows fast. SPARQL Endpoint becomes main preferences to access data because it is a flexible way to interact with Web of Data by formulating query like SQL in traditional database. Additionally, it returns query answer in several formats, such as XML and JSON  which are widely used as data exchange standard in various applications. This situation has attracted people to aggregate data from multiple SPARQL Endpoints akin to conventional distributed databases. For instance, NeuroWiki\footnote{\url{http://neurowiki.alleninstitute.org/}} collects data from multiple life science RDF store by utilizing LDIF framework \cite{www12schultz} and RKBexplorer\footnote{\url{ http://www.rkbexplorer.com/explorer/}} gathers research publication information from more than 20 datasets under \url{rkbexplorer.com} domain \cite{DBLP:conf/semweb/MillardGSS10}.

\begin{figure}
  \centering
	\includegraphics[width=.97\linewidth]{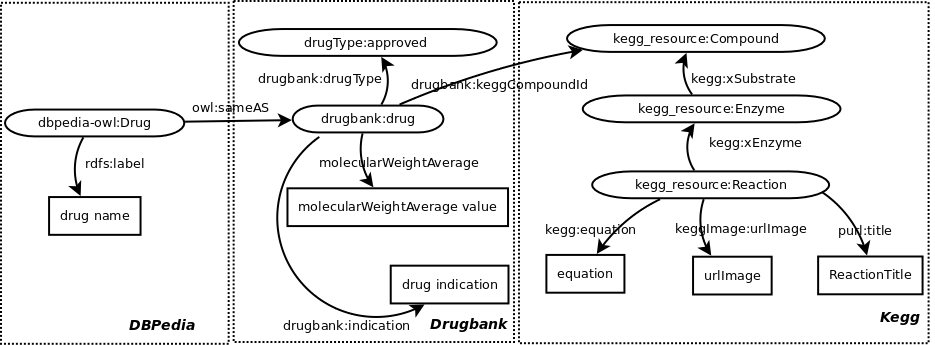}	
	\caption{Example Data relation located at DBpedia, Drugbank and Kegg Dataset}
	\label{fig:dbpediadrugbankkegg}
\end{figure}

Querying data in the Web of Data context is more challenging than collecting information in traditional distributed databases, as Web of Data has no global schema, typically offering heterogeneity in terms of vocabularies~\cite{freitas2012querying}. The publishers use their own vocabulary to produce their data, therefore the consumer should understand the layout of the dataset before querying it. In contrast, traditional databases provide global schema, allowing consumers to request data in a straightforward manner.
To deal with the lack of a global schema,  querying data from multiple sources could be solved by link establishing among datasets~\cite{Heath2011}. The publisher only needs to generate a link from his dataset to other dataset by doing entity matching among multiple datasources. Entity matching is the process of connecting two entities located in different datasets, related to each other. SILK \cite{Jentzsch2010} and LIMES \cite{Ngomo_Auer_2011} are two tools allowing to generate links semi-automatically. The set of interlinked Web of Data datasets creates Linked Data\footnote{\url{http://www.w3.org/standards/semanticweb/data}}. According to  Linked Open Data (LOD) cloud statistics\footnote{\url{http://lod-cloud.net/state/}}, as of September 2011, 89.83 \% of datasets has more than 1000 links to other datasets. Those links are beneficial to aggregate data from multiple datasets.  Consider, for example, all drug information in DBpedia (\url{http://dbpedia.org}) can be connected with drugs in Drugbank (\url{http://www4.wiwiss.fu-berlin.de/drugbank/)} by the \textit{owl:sameAs} relation; which is an identity link that joins two entities having the same identity. This may result in easy querying data amongst multiple SPARQL Endpoints. For example, if we search data about a drug, we can collect data from \textit{Drugbank}\footnote{\url{http://www4.wiwiss.fu-berlin.de/drugbank/sparql}}, \textit{DBPedia}\footnote{\url{http://dbpedia.org/sparql}} and \textit{Kegg}\footnote{\url{http://s4.semanticscience.org:16036/sparql}} SPARQL Endpoints as shown in Figure~\ref{fig:dbpediadrugbankkegg}. The oval in the Figure~\ref{fig:dbpediadrugbankkegg} represents entity whereas the box denotes property value. Querying data from three datasets produces drug information such as its indication and compounds. We obtain drug information and its indication from Drugbank and DBpedia by utilizing \textit{owl:sameAs}. The \textit{owl:sameAs} link is a shortcut to query data from two dataset without knowing Drugbank and DBpedia schemas. 

In this study, we will focus primarily on federation over SPARQL Endpoint infrastructure, as the LOD cloud statistics reports that 68.14\%  of the RDF repositories are equipped with SPARQL Endpoints. Other infrastructures that use query languages such as RQL\footnote{\url{http://139.91.183.30:9090/RDF/RQL/}}, RDQL\footnote{\url{http://www.w3.org/Submission/RDQL/}}, SeRQL\footnote{\url{http://www.w3.org/2001/sw/wiki/SeRQL}} are beyond the scope of this study. 
Aside from giving an overview of querying over SPARQL Endpoints, we will compare existing federation frameworks based on their platform, infrastructure properties, query processing strategies, etc.---to chose any framework for small and large-scale systems.  Further, we highlight shortcomings in the current federation frameworks that could open an avenue in the research of federated queries. In addition, we propose several features that should be added for further federation framework development.

We initially present related survey and evaluation of Federation query at Section~\ref{sec-rel}. A real-world use case of to motivate query federation in the Health Care and Life Sciences (HCLS) domain is presented in Section~\ref{motivatingExample}. Section~\ref{dataintegration}  introduces the concept of data integration in Linked Data that gives the necessary foundation of infrastructure for querying over Linked Data. We provide an overview of federation architectures and detail the phase of querying over SPARQL Endpoint in Section~\ref{federation} and in Section~\ref{fed-exist} introduce existing  federation frameworks, supporting either SPARQL 1.0 or 1.1\footnote{\url{http://www.w3.org/TR/sparql11-query/}}. We also categorize them based on their architecture and querying process and investigate features that should be added in the existing frameworks. Finally, we discover challenges that should be considered in the future development of federation query in Section~\ref{Challenges}. We conclude our finding in Section~\ref{conclusion}.

\section{Related Works}
\label{sec-rel}
More general investigations w.r.t. querying Linked Data have been performed elsewhere~\cite{DBIS:BGHS12,Gorlitz_federateddata,Ladwig:2010:LDQ:1940281.1940311}. \cite{DBIS:BGHS12} mentioned nine myths and five challenges arising in the Federation over Linked Data. Based on their observation, they suggested to consider Linked Data as a service not as distributed data. \cite{Gorlitz_federateddata} explained the Federation query infrastructure, whereas \cite{Ladwig:2010:LDQ:1940281.1940311} focused on the basics of federation query processing strategy. 

A number of studies \cite{MontoyaVCRA12ISWC2012,DBLP:journals/corr/abs-1210-5403} compare federation frameworks by evaluating their performance. \cite{MontoyaVCRA12ISWC2012} tests federation frameworks by using FedBech \cite{SchmidtGHLST11} in various networking environment and data distribution.  Similar to \cite{MontoyaVCRA12ISWC2012}, \cite{DBLP:journals/corr/abs-1210-5403} conducts an experiment in the FedBench to evaluate federation frameworks on large scale Life Science datasets. In this survey, we investigate and compare more existing Federation over SPARQL Endpoints frameworks based on their strategy such as source selection and execution plan.

\section{Motivating Example}
\label{motivatingExample}
The HCLS domain advocated Linked Data from its early days, and at present a considerable portion of the Linked Data cloud is comprised of datasets from Linked Data for Life Sciences (LD4LS) domain~\cite{HasnainOEDW2012}. LD4LS currently comprises multiple datasets from certain HCLS projects, namely \textit{bio2rdf}\footnote{\url{http://bio2rdf.org/}}, the Health Care and \textit{Life Sciences Knowledge base}\footnote{\url{http://www.w3.org/TR/hcls-kb/}} (HCLS Kb), \textit{linkedlifedata}\footnote{\url{ http://linkedlifedata.com/}}, Linked Open Drug Data effort\footnote{\url{http://www.w3.org/wiki/HCLSIG/LODD}} and the \textit{Neurocommons}\footnote{\url{http://neurocommons.org/page/Main_Page}}. These efforts have been derived and are still motivated in biomedical facilities in the recent years, partially caused by the decrease in price for acquiring large datasets such as genomics sequences and the trend towards personalized medicine, pharmacogenomics and integrative bioinformatics, to access and query life sciences data. \cite{Hernandez:2004:IBS:1031570.1031583} described the high demand of biological datasets integration to help life science researcher. Although the publication of datasets as RDF is a significant milestone to achieve the ability to query these healthcare and other biological datasets, to this date, it is a big deal to enable a query-able Web of HCLS data. 

To achieve the ability for assembling queries encompassing multiple graphs hosted at various places, it is therefore critically necessary that vocabularies and ontologies are reused~\cite{polleres2010semantic}. This can be achieved either by ensuring that the multiple datasets make use of the same vocabularies and namespaces or assemble a federated query over multiple datasets to retrieve a meaningful information. In order to understand need of federation, consider two SPARQL Endpoints DrugBank and Kegg (Figure~\ref{fig:dbpediadrugbankkegg}) that publish information regarding drugs and compounds. Both the datasets have different useful information about the same concepts. In order to find the answer of the question \textit{Find the Chemical equations and Reaction titles of reactions related to only those drugs which are approved along with average Molecular Weight}, one should send the query to two SPARQL Endpoints Drugbank and Kegg. Kegg contains two concepts namely Chemical equations and Reaction title whereas the information like average Molecular Weight and ``approved drugs'' is present at Drugbank endpoint. This complex and related information can be retrieved using Query~\ref{lifescienceexample} that should be federated to Kegg and Drugbank endpoints.

\section{Infrastructure for Querying Linked Data}
\label{dataintegration}
\begin{figure}
  \centering
	\includegraphics[width=.6\linewidth]{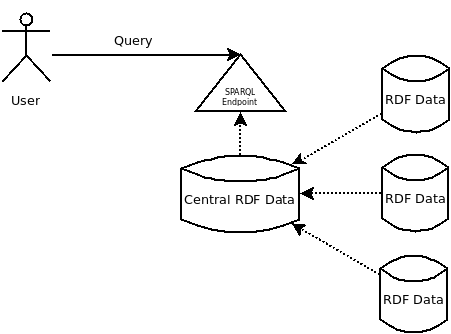}	
	\caption{Central Repository}
	\label{fig:centralrepo}
\end{figure}

Based on data source location, the infrastructure for querying Linked Data can be divided into two categories, namely 1) central repository and distributed repositories. Central repository has similar characteristic as of data warehousing in traditional databases, where the data is collected in advance in a single repository before query processing (Figure~\ref{fig:centralrepo}). Sindice\cite{Tummarello2007SWO17851621785203} is an example of a central repository, which crawls data, indexes it and provides APIs and a SPARQL Endpoint for accessing the data. The efficiency of query time is one advantage because data has already been placed at one location. The single data location in this scenario offers following benefits : no network communication and no source selection required. However, due to the frequently changing data source, the data synchronization could be a problem~\cite{Umbrich-ISWC2012}.  Furthermore, the data storage needs a lot of space in order to keep data in one place. Not only consuming more space, but this approach is resource intensive in regard to processing large scale data.

\lstset{language=SPARQL, caption=Data Integration Linked Data Example in Life Science Domain , label=lifescienceexample,frame=single,breaklines=true,basicstyle=\small}
\begin{lstlisting}[float]
PREFIX drugbank: <http://www4.wiwiss.fu-berlin.de/drugbank/resource/drugbank/> 
PREFIX drugType: <http://www4.wiwiss.fu-berlin.de/drugbank/resource/drugtype/>
PREFIX kegg:<http://bio2rdf.org/kegg_vocabulary:>
PREFIX keggImage:<http://bio2rdf.org/ns/bio2rdf#>
PREFIX purl:<http://purl.org/dc/elements/1.1/>

SELECT distinct ?drug  ?drugtype  ?compound ?molecularWeightAverage ?ReactionTitle ?ChemicalEquation 

WHERE {
   ?drug drugbank:drugType drugType:approved .
   ?drug drugbank:keggCompoundId ?compound .
   ?drug drugbank:molecularWeightAverage ?molecularWeightAverage.
   ?enzyme kegg:xSubstrate ?compound .
   ?Chemicalreaction kegg:xEnzyme ?enzyme .
   ?Chemicalreaction kegg:equation ?ChemicalEquation . 
   ?Chemicalreaction keggImage:urlImage "http://www.genome.jp/Fig/reaction_small/R05248.gif".
   ?Chemicalreaction purl:title ?ReactionTitle
}
\end{lstlisting}

\begin{figure}
  \centering
	\includegraphics[width=.6\linewidth]{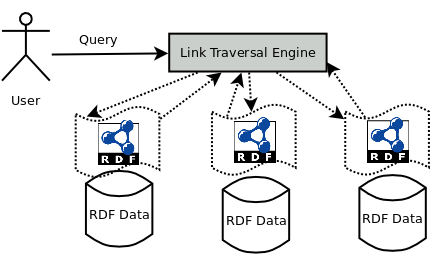}	
	\caption{Link Traversal}
	\label{fig:linktraversal}
\end{figure}

As opposed to central repository, querying Linked Data in distributed repositories environment does not need crawling data beforehand. We group distributed repositories in two systems: Link Traversal and Federation. Discovering data by following HTTP URIs is the basic idea in the link traversal system. As illustrated in Figure~\ref{fig:linktraversal}, without any data knowledge, relevant data sources are detected during runtime execution~\cite{Hartig:2011:ZQP:2008892.2008906}. Link traversal provides high freshness of the data since the data is directly accessed from data source.
The query execution  is initially from one single triple pattern as starting point. Determining the starting point is a vital task in this system because it influences the flow process of whole of query execution. The wrong starting point decision can increase intermediate results, as a result the bandwidth usage goes up. Another drawback of this system is the limitation of type executable query such as query pattern consisting unbound predicate. 

\begin{figure}
  \centering
	\includegraphics[width=.7\linewidth]{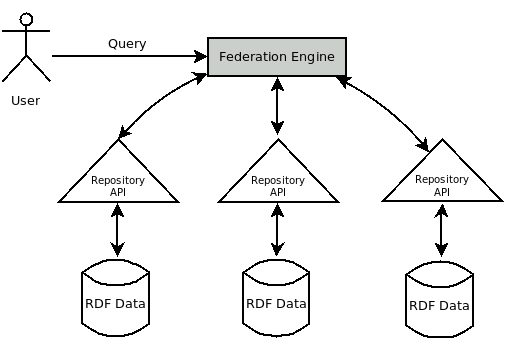}	
	\caption{Federation over Single Repositories}
	\label{fig:fedSR}
\end{figure}

\begin{figure}
  \centering
	\includegraphics[width=.7\linewidth]{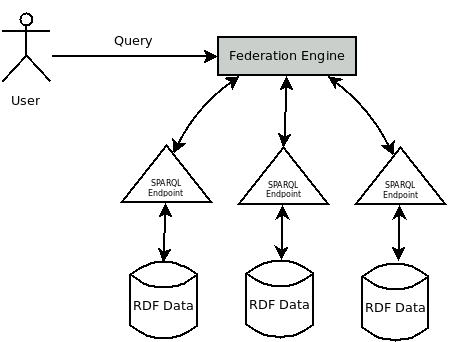}	
	\caption{Federation over SPARQL Endpoint}
	\label{fig:fedSE}
\end{figure}

Federation uses a query mediator to transform a user query into several sub queries and generates results from the integrated data sources. As the data sources need not be collected in one repository, the data is more up-to-date than central repository's result, but query processing time takes longer than the case of central repository infrastructure. The system only invests little resource such space and time because of no earlier crawling data phase. In contrast to central repository, overhead communication between mediator and data source often occurs in the federation system since source selection is required during query execution. There are two kinds of federation frameworks: federation over single repositories (Figure~\ref{fig:fedSR}) and federation over SPARQL Endpoints (Figure~\ref{fig:fedSE}). In the federation over single repositories such as Sesame Sail Federation\footnote{\url{http://www.openrdf.org/alibaba.jsp}}, the federation query interface delivers sub queries through native API of its repository. Nevertheless, not all repositories support this API. Furthermore, the loading data from RDF dump to repository must be done in advance before query transmitted, thus we do not take account this distribution in our observation. The other distributed query type requires SPARQL Endpoint as bridge between federation layer and datasource (we will describe this type of federation in section \ref{fed-exist}).  Most of the federation frameworks are compatible with this type, since RDF store are generally equipped with SPARQL Endpoint. Several mediators sometimes~\cite{Langegger:2008:SWM:1789394.1789441,acosta2011eaqptffose} offer wrapper supporting other data format like CSV and XML.  Hence, in this study we only consider to survey federation over SPARQL Endpoints and federation supporting wrapper to access non RDF graph that could be accessed like SPARQL Endpoints.

\section{Federation over SPARQL Endpoints Basic Approach and Technique}
\label{federation}
This section provides an overview of architecture and basic mechanisms that can be used in any kind of federation over SPARQL Endpoints framework.
\subsection{Architecture of Federation over SPARQL Endpoints}
\label{fed-Arch}
\begin{figure}
  \centering
	\includegraphics[width=.8\linewidth]{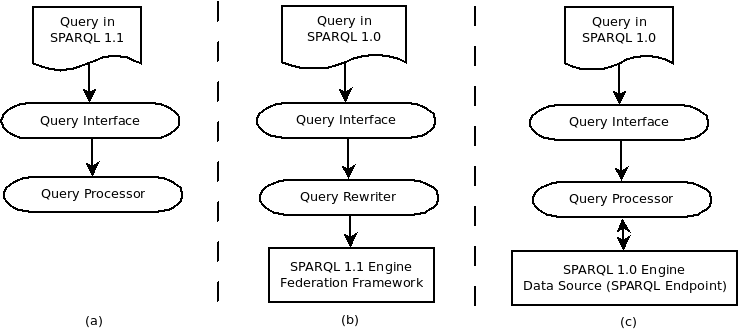}	
	\caption{Architecture of Federation over SPARQL Endpoint}
	\label{fig:architecture}
\end{figure}

SPARQL 1.1 is designed to tackle limitations of the SPARQL 1.0, including updates operations, aggregates, or federation query support. As of this writing, not all query engines support SPARQL 1.1.  Therefore, we discuss the federation frameworks that support either SPARQL 1.0 or SPARQL 1.1. There are three kinds of architecture of federation over SPARQL Endpoints (Figure~\ref{fig:architecture}) namely a) framework has capability to execute SPARQL 1.1 query, b) framework accepts SPARQL 1.0, then it rewrites query to SPARQL 1.1 syntax before passing it to SPARQL 1.1 engine and c) framework handles SPARQL 1.0 and processes the query in several phases by interacting with SPARQL 1.0 engine of each SPARQL Endpoints.
Those system that has already supported SPARQL 1.1 allow user to execute query federation over SPARQL Endpoints by using \textit{SERVICE} operator (Figure~\ref{fig:architecture}.a). The query processor sends each sub query to defined SPARQL Endpoint and join the result from SPARQL Endpoint. Basically, SPARQL 1.0 allows us to query data from remote data sources, however it does not retrieve specified remote SPARQL Endpoints. As described in the Query~\ref{FROMremote}., it only fetches remote graphs or graphs with the name in a local store.

At present, the SPARQL 1.1 is the simplest solution to yield data from multiple sources. The W3C recommendation of SPARQL 1.1 formalizes rule to query in multiple SPARQL Endpoints by using \textit{SERVICE} operator. However, users must have prior knowledge regarding the data location before writing a query because the data location must be mentioned explicitly. As seen in the Query~\ref{owlsameasSERVICE}, the Drugbank and DBPedia SPARQL Endpoints are mentioned after SERVICE operator to obtain the list of drugs and their associated diseases. In order to assist users in term of data source address, it allows us to define a list of SPARQL Endpoints as data beforehand and attach it as variable in the SPARQL query. Besides \textit{SERVICE}, SPARQL 1.1 also introduces \textit{VALUES} as one of SPARQL Federation extension. It can reduce the intermediate results during query execution by giving constrains from the previous query to the next query. 

\lstset{language=SPARQL, caption=Example of Federated SPARQL Query in the SPARQL 1.0, label=FROMremote,frame=single,breaklines=true,basicstyle=\small}
\begin{lstlisting}[float]
SELECT ?drugname ?indication 
WHERE {
FROM <http://localhost/dbpedia.rdf>
{
	?drug a dbpedia-owl:Drug .
	?drug rdfs:label ?drugname .
	?drug owl:sameAs ?drugbank .
}
FROM <http://localhost/drugbank.rdf>
{
	?drugbank drugbank:indication ?indication .
}
}
\end{lstlisting}

\lstset{language=SPARQL, caption=Example of Federated SPARQL Query in the SPARQL 1.1, label=owlsameasSERVICE,frame=single,breaklines=true,basicstyle=\small}
\begin{lstlisting}[float]
SELECT ?drugname ?indication 
WHERE {
SERVICE <http://dbpedia.org/sparql>
{
	?drug a dbpedia-owl:Drug .
	?drug rdfs:label ?drugname .
	?drug owl:sameAs ?drugbank .
}
SERVICE <http://www4.wiwiss.fu-berlin.de/drugbank/sparql>
{
?drugbank drugbank:indication ?indication .
}
}
\end{lstlisting}

The lack of knowledge data information is a main problem to execute federation query on single RDF store. Thus, several efforts have been introduced to address that issue (Figure~\ref{fig:architecture}.b). The user can write a query blindly without knowing the data location. These federation models can executes Query~\ref{owlsameasSERVICE} or Query~\ref{FROMremote} without a SPARQL Endpoint declared. By removing SERVICE or FROM keywords, those two queries can be replaced by Query~\ref{owlsameas}. These framework architectures provide an interface to translate query from SPARQL 1.0 to SPARQL 1.1 format. The core part of this interface is query rewriting component. After parsing and decomposing the query, this component adds destination address of this query by inserting \textit{SERVICE} operators in each sub query. Further on, the result of query rewriter will be executed by internal SPARQL 1.1 processor system.

\lstset{language=SPARQL, caption=Example of Federation SPARQL Query in the SPARQL 1.0 without SPARQL Endpoint specified , label=owlsameas,frame=single,breaklines=true,basicstyle=\small}
\begin{lstlisting}[float]
SELECT ?drugname ?indication
WHERE {
?drug a dbpedia-owl:Drug .
?drug rdfs:label ?drugname .
?drug owl:sameAs ?drugbank .
?drugbank drugbank:indication ?indication .
}
\end{lstlisting}

Since not all current SPARQL Endpoints can handle SPARQL 1.1 query, several systems (Figure~\ref{fig:architecture}.c)  developed query execution processor to execute federation SPARQL query in SPARQL 1.0 format. The processor has responsibility to manage query processing such as maintain data catalogue, determine relevant sources, plan the query execution and join all results after retrieving data from SPARQL Endpoints.

\subsection{Basic Steps of Federation over SPARQL Endpoints}
\label{fed-steps}
\begin{figure}
  \centering
	\includegraphics[width=.5\linewidth]{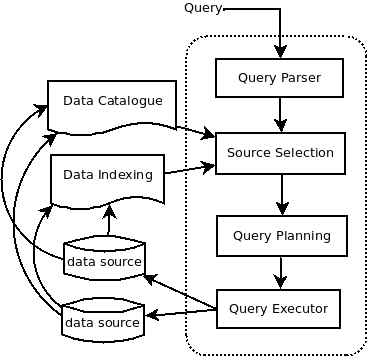}	
	\caption{Federated SPARQL Query Process}
	\label{fig:process}
\end{figure}

A mediator holds important role to manage an incoming query from user, deliver query to each data source and gives back the result to the user. We explain some important concept around of federation frameworks for executing a query that is depicted at Figure~\ref{fig:process}. in the following section. 
\subsubsection{Query parser}\hfill \\
In this initial phase, SPARQL query is transformed to internal pattern which can be in the list of Basic Graph Pattern (BGP), abstract syntax tree~\cite{Gorlitz2011}, or other formats. BGP is a set of triple patten where one triple consists of subject, predicate and object. One or more variables could be in subject, predicate and object position. The outcome of the parsing step is useful for later steps, particularly in query optimization. 

\subsubsection{Source Selection}\hfill \\
Instead of sending every piece of query to all data sources, the mediator should determine the relevant source of sub query carefully. The simple query might not become a big issue when delivering query to all destinations, however, regardless the capability of source for answering a query, the transmission of complex query with many intermediate results could lead expensive communication cost. Choosing a relevant source for a query could be done in several following methods: 
\begin{itemize}
\item ASK SPARQL Query \hfill \\
ASK SPARQL query returns boolean value that decides whether the query can be answered by the SPARQL Endpoint or not. For instance, to solve Query~\ref{lifescienceexample}, the mediator will send Query~\ref{AskQuery} to seek sources that can answer sub query    \textit{?drug drugbank:molecularWeightAverage ?molecularWeightAverage}. By sending ASK query, the bandwidth usage of query execution can be reduced significantly because a query is only transferred to the Endpoint responding true value. Furthermore, it can detect the changing data easily in runtime. The limitation of ASK is that it is  only a binary decision and it can not detect redundancy data among data sources, therefore~\cite{Hose:2012:TBR:2237867.2237869} extended the ASK operation during data source selection by including a sketch: estimation number of result and summary of the result. In addition, this methods is only suitable for few number of source participated since it takes longer time for waiting each SPARQL Endpoint answering ASK query.

\lstset{language=SPARQL, caption=Example ASK SPARQL Query to Select Relevant Source, label=AskQuery,frame=single,breaklines=true,basicstyle=\small}
\begin{lstlisting}[float]
PREFIX drugbank: <http://www4.wiwiss.fu-berlin.de/drugbank/resource/drugbank/> 
ASK {
   ?drug drugbank:molecularWeightAverage ?molecularWeightAverage .
}
\end{lstlisting}

\item Data catalogue \hfill \\
By looking up data catalogue of a dataset, the mediator can predict the suited sources for a query. \cite{Langegger:2008:SWM:1789394.1789441,Gorlitz2011} utilize VoID~\cite{Alexander09describinglinked} as data catalogue. LOD cloud statistic states that 32.2\% of LOD datasets provide dataset descriptions expressed in VoID. VoID expresses the metadata of the dataset and its relation with other datasets. It is useful for registering new data source in the SemWIQ~\cite{Langegger:2008:SWM:1789394.1789441}, whereas it is used to estimate the SPARQL query pattern cardinalities~\cite{5767868} in SPLENDID~\cite{Gorlitz2011}.  \cite{Langegger:2008:SWM:1789394.1789441,Gorlitz2011} also employs RDFStats~\cite{Langegger:2009:RER:1674635.1674691} which is built on SCOVO~\cite{Hausenblas:2009:SUS:1561533.1561592}. RDFStats comprises instance statistics and histogram of class, property and value type. Both of VoID and RDFStats can be used independently in any case from those frameworks. Service Description~\footnote{\url{http://www.w3.org/TR/sparql11-service-description/}} describes the data availability from SPARQL Endpoint, data statistic and the restriction of query pattern. \cite{aderis_2010} constructs a data catalogue containing list of predicates during setup and querying phase. In general, data catalogue only provides a list of predicate for each source since the number of predicates is less than number of subjects and objects. Based on the catalogue, mediator commonly does pre computing statistic that is needed for query optimization. The data catalogue can be updated during query execution especially for frequently changing data. The freshness of data catalogue has impact on the accuracy of source selection, but it consumes much bandwidth during updating  process. Furthermore, the SPARQL query delivered for updating data catalogue is expensive operation which might be refused by SPARQL Endpoint.

\item Data Indexing \hfill \\
According to LOD cloud statistic, 63\% of data sources do not expose their data catalogues. In order to overcome the lack of data catalogue, the data indexing process could be done before or during query execution. \cite{Harth:2010:DSO:1772690.1772733} introduced indexing method by applying QTrees~\cite{Hose:2005:PRQ:1783738.1783759}. This indexing consists of description of instances and schema that could assist mediator to determine relevant source for a query. In general, the drawback of data indexing is needs more storage than data catalogue as it usually indexes every triple. In order to deal with space problem,
\cite{Basca_Bernstein_2010} detects the relevant query sources by obtaining information from either Search Engine such as Sindice\footnote{\url{http://sindice.com/}} or Web Directory after query parsing step. As a result, a summary of ontological prefixes, predicates and classes are indexed for query execution process. However, network latency issue during execution arises because it relies on third party indexing.

\item Caching \hfill \\
Most of the federation frameworks load statistical information during initialization phase because they want to reduce the bandwidth consumption during runtime. However, the source selection process could be not accurate, if the data is frequently changing. To tackle this shortcoming, the cache during query execution can be employed for later query execution. The information stored at a cache is likely similar to data catalogue. To decrease the communication cost, the mediator does not deliver new query, but the updating statistic data source information based on the result of sub query answers from SPARQL Endpoint.
\end{itemize}

All aforementioned source selection strategies can be applied to dynamic data such as data stream as long as it is always updated. As updating task consumes much bandwidth, we should have a mechanism that only requests for frequently changing data as proposed by \cite{Umbrich-ISWC2012} or only updates data information periodically.

%

\subsubsection{Query Planning and Execution}\hfill \\
In the second phase of query processing, the query mediator decomposes the query and builds multiple sub queries according to the result of source selection. One sub query can be delivered to multiple sources if many sources can contribute the answer. Based on selection source stage, mediator usually does matching pattern. 

The construction of sub query is also considered before query transmission process. One single triple pattern can match either one source or multiple sources. To reduce repetition of one triple query sending to multiple source, a triple can be grouped with other triples in one sub query. The grouping query triples can also minimize the intermediate join process. \cite{FedX2011} and~\cite{WoDQA2012} proposed Exclusive Group scheme to cluster related triple patterns in one sub query. After building sub queries for each data source, the mediator arranges the order of sub queries in various combination of execution plans. Hence, federation framework employs statistical information to compute cost execution of each plan. Later on, the execution plan with the lowest cost will be chosen and executed in the following strategies:  

\begin{itemize}
\item Nested Loop Join \hfill \\
Nested loop join is not optimal solution for complex query since every previous scanning results will be joined to next result. 

\item Bind join \hfill \\
To improve the nested loop join, bind join, firstly introduced by \cite{Haas:1997:OQA:645923.670995}, passes the intermediate result to become filter for next query. As a result, the transfer cost can be minimized, but the query runtime takes longer because the query mediator have to wait for the complete answer of previous query.

\item Hash join \hfill \\
Hash join is implemented in~\cite{Gorlitz2011} which joins all intermediate result locally after submitting sub queries in parallel.  This join could boost the runtime performance for small intermediate result. However, the transmission cost will be higher if the intermediate result is large.

\end{itemize}

\section{The Existing Federation over SPARQL Endpoints Frameworks}
\label{fed-exist}
This section presents the insight on existing federation over SPARQL Endpoint based on their architectures. To give better explanation, they will be classified based on their features. Ultimately, we propose several features that should be considered for next development.

\subsection{The  Existing Federation over SPARQL Endpoints Frameworks Based on Their Architecture}
As described in the section~\ref{fed-Arch}, three federation architectures categories have been developed recently. We explain the existing federation  based on those categories in this section.
\subsubsection{Frameworks Support SPARQL 1.1 Federation Extension}  \hfill \\
As of this writing, several RDF store systems have been able to process federation query, but not all of them support \textit{VALUES} keyword. Instead of handling \textit{VALUES}, a number of frameworks has supported  \textit{BINDING} which is also addressed to reduce the size of intermediate results. The list of existing frameworks supporting SPARQL 1.1 presents at Table \ref{tab:currentdatafed11}.
\begin{table}
\caption{The Existing Frameworks Support SPARQL 1.1 Federation Extension.}
\label{tab:currentfedservice}
\begin{center}
\begin{tabular}{|c|c|c|c|c|}
\hline 
Framework & Platform & SERVICE & BINDINGS & VALUES \\ 
\hline 
ARQ & Jena & \tickYes & \tickNo & \tickYes \\ 
\hline 
SPARQL-FED & Virtuoso & \tickYes & \tickNo & \tickYes\\ 
\hline 
Sesame & Sesame & \tickYes & \tickYes & \tickYes\\ 
\hline 
SPARQL-DQP &  OGSA-DAI and OGSA-DQP & \tickYes & \tickYes & \tickNo\\ 
\hline 
\end{tabular} 
\end{center}
\end{table}

\begin{description}

\item[ARQ] \hfill \\
ARQ\footnote{\url{http://jena.apache.org/documentation/query/index.html}}, a query engine processor for Jena, has supported federated query by providing \textit{SERVICE} and \textit{VALUES} operator. ARQ implements nested loop join to gather retrieved result from multiple SPARQL Endpoints. In term of security, the credential value to connect ARQ service must be initialized in the pre-configuration\footnote{\url{http://jena.hpl.hp.com/Service#}}.

\item[Sesame] \hfill \\
Previously, Sesame already supported federation SPARQL query by using SAIL AliBaba extension\footnote{\url{http://www.openrdf.org/doc/alibaba/2.0-alpha2/alibaba-sail-federation/index.html}} at 2009, but it can not execute SPARQL 1.1. Instead, it integrates multiple datasets into a virtual single repository to execute federated query in SPARQL 1.0. It can execute federation SPARQL query either RDF dump or SPARQL Endpoint by using its API. The data source must be registered in advance during setup phase. The simple configuration file only containing the list of SPARQL Endpoint address can cause poor performance since it sends query to all data source without source selection. In order to optimize the query execution, it offers additional features in the configuration file namely predicate and subject prefixes owned by one dataset. According to the configuration, it can do prefix matching to predict the relevant source for a sub query. The join ordering is decided by calculating the size of basic graph pattern. The new version of sesame (2.7)\footnote{\url{http://www.openrdf.org/index.jsp}} is able to handle SPARQL 1.1 which provides Federation extension features including \textit{SERVICE} and \textit{VALUES} operator.

\item[SPARQL-FED] \hfill \\
Virtuoso 6.1 allows to execute SPARQL queries to remote SPARQL Endpoint through SPARQL-FED\footnote{\url{http://www.openlinksw.com/dataspace/dav/wiki/Main/VirtSparqlCxml}}. The remote SPARQL Endpoint must be declared after SERVICE operator.

\item[SPARQL-DQP]\hfill \\
It is built on top of the OGSA-DAI~\cite{ogsadai3-whats-w} and OGSA-DQP~\cite{Lynden:2009:DIO:1460947.1461353} infrastructures. It transforms incoming SPARQL query to SQL, as it implements SQL optimization techniques to generate and optimize query plans. The optimization strategy is based on OGSA-DQP algorithm which does not need any statistic information from data sources. No SPARQL Endpoint registration is required because the SPARQL Endpoint must be written in query. The OGSA-DAI manages a parallel hash join algorithm to reorder query execution plan.
\end{description}
\subsubsection{Frameworks Supports Federation over SPARQL Endpoint, build on top of SPARQL 1.0}  \hfill \\
In order to overcome data location knowledge, several federation frameworks have been developed recently without specifying SPARQL Endpoint address in the query. The federation framework acts as mediator~\cite{springerlink:10.1007/978-3-642-23032-54} that transfers SPARQL query from user to multiple data source either RDF repository or SPARQL Endpoints. Before delivering query to destination source, it  breaks down a query into sub queries and selects the destination of each sub query. In the end, the mediator must join the retrieved result from the SPARQL Endpoints. Following are overview of the current of federation frameworks and summarized in Table~\ref{tab:currentdatafed10}.
\begin{description}
\item[DARQ] \hfill \\
DARQ (Distributed ARQ)~\cite{DARQQuilitz2008} is an extension of ARQ which provides transparent query to access multiple distributed endpoint by adopting query mediator. The service description which consists of data description and statistical information has to declare in advance before query processing to decide where a sub query should go. According to the list of predicates in the service description, it re-writes the query, creates sub query and design the query planning execution. The query planning is based on estimated cardinality cost. DARQ implements two join strategies : Nested Loop Join and Bind Join.

\item[Splendid] \hfill \\
Splendid~\cite{Gorlitz2011} extends Sesame which employs VoID as data catalogue. The VoID of dataset is loaded when the system started then ASK SPARQL query is submitted to each dataset for verification. Once the query is arrived, the system builds sub queries and join order for optimization. Based on the statistical information, the bushy tree execution plan is generated by using dynamic programming \cite{Selinger:1979:APS:582095.582099}. Similar to DARQ, it computes join cost based on cardinality estimation. It provides two join types: hash join and bind join to merge result locally.

\item[FedX] \hfill \\
FedX~\cite{FedX2011} is also developed on top of the Sesame framework. It is able to run queries over either Sesame repositories or SPARQL Endpoints. During initial phase, it loads the list of data sources without its statistical information. The source selection is done by sending SPARQL ASK queries. The result of a SPARQL ASK query is stored in a cache to reduce communication for successive query. Intermediate result size is minimized by a rule based join optimizer according to cost estimation. It implements Exclusive Groups to cluster related patterns for one relevant data source. Beside grouping patterns, it also groups related mapping by using single SPARQL UNION query. Those strategies can decrease the number of query transmission and eventually, it reduces the size of intermediate results. As complementary, it came with Information Workbench for demonstrating the federated approach to query processing with FedX. 

\item[ADERIS] \hfill \\
ADERIS (Adaptive Distributed Endpoint RDF Integration System)~\cite{aderis_2010} fetches the list of predicates provided by data source during setup stage. The predicate list can be used to decide destination source for each sub query pattern. During query execution, it constructs predicate tables to be added in query plan. One predicate table belongs to one sub query pattern. The predicate table consists of two columns : subject and object which is filled from intermediate results. Once two predicate tables have completed, the local joining will be started by using nested loop join algorithm. The predicate table will be deleted after query is processed. ADERIS is suitable for data source who does not expose data catalogue, but it only handles limited query patterns such as UNION and OPTIONAL. The simple GUI for configuration and query execution are provided by ADERIS.

\item[Avalanche] \hfill \\
Avalanche~\cite{Basca_Bernstein_2010} does not maintenance the data source registration as its data source participant depends on third party such as search engine and web directory. Apart from that, it also stores set of prefixes and schemas to special endpoints. The statistic of data source is always up to date since it always requests the related data source statistic to search engine after query parsing. To detect the data source that contributes to answer a sub query, it calculates the cardinality of each unbound variables. The combinations of sub queries are constructed by utilizing best first search approach.  All sub queries are executed in parallel process. To reduce the query response time, it only retrieves first K results from SPARQL Endpoint.

\item[GDS] \hfill \\
Graph Distributed SPARQL (GDS)~\cite{wang2011querying} overcomes the limitation of their previous work~\cite{conf:aswc:WangTD11} which can not handle multiple graphs. It is developed on top of Jena platform by implementing Minimum Spanning Tree (MST) algorithm and enhancing BGP representation. Based on Service description, MST graph is generated by exploiting Kruskal algorithm  which aims to estimates the minimum set of triple patterns evaluation and execution order. The query planning execution can be done by either semi join or bind join which is assisted by cache to reduce traffic cost.
\item[Distributed SPARQL] \hfill \\
In contrast to the above frameworks, users must declare the SPARQL Endpoint explicitly in the SPARQL query at Distributed SPARQL~\cite{DBLP:conf:semweb:ZemanekS08}. Since it is developed for SPARQL 1.0 user, the SPARQL Endpoint address is mentioned after \textit{FROM NAMED} clause. Consequently, this framework does not require any data catalogue to execute a query. As part of Networked Graphs~\cite{Schenk:2008:NGD:1367497.1367577}, it is also built on the top of Sesame. To minimize number of transmission query during execution, it applies distributed semi join in the query planning.
\end{description}
\begin{table}
\caption{The Existing Frameworks Supports Federation over SPARQL Endpoints without reformulating query to SPARQL 1.1.}
\label{tab:currentdatafed10}
\begin{tabular}{|p{1.6cm}|p{2cm}|p{1.5cm}|p{2cm}|p{1cm}|p{2cm}|p{1.4cm}|p{0.6cm}|}
\hline 
Framework & Catalogue & Platform & Source Selection & Cache  & Query Execution & Source Tracking & GUI\\ 
\hline 
DARQ & Service Description & Jena & 
Statistic of Predicate
 & \tickYes  & Bind Join or Nested Loop Join  & Static & \tickNo \\ 
\hline 
ADERIS & Predicate List during setup phase & \tickNo & Predicate List & \tickNo  & Nested Loop Join &  Static & \tickYes \\ 
\hline 
FedX & \tickNo & Sesame & ASK & \tickYes  & 
Bind Join parallelization & Dynamic & \tickYes \\ 
\hline 
Splendid & VoID & Sesame & Statistic + ASK & \tickNo & Bind Join or Hash Join  & Static & \tickNo \\ 
\hline 
GDS & Service Description & Jena & Statistic of Predicate & \tickYes & Bind Join or Semi Join & Dynamic & \tickNo \\
\hline 
Avalanche & Search Engine & Avalanche & Statistic of predicates and ontologies & \tickYes  & Bind join & Dynamic & \tickNo \\ 
\hline 
Distributed SPARQL & \tickNo & Sesame & \tickNo & \tickNo & Bind join & \tickNo & \tickNo \\ 
\hline
\end{tabular} 
\end{table}
\subsubsection{Frameworks Supports Federation over SPARQL Endpoint, build on top of SPARQL 1.1}  \hfill \\
A number of frameworks were developed to accepts SPARQL query federation in SPARQL 1.0 format, but they are built on top of SPARQL query engine that support SPARQL 1.1 (Table~\ref{tab:currentdatafed11}.)

\begin{description}
\item[ANAPSID] \hfill \\
ANAPSID~\cite{acosta2011eaqptffose} is a framework  to manage query execution with respect to data availability and runtime condition for SPARQL 1.1 federation. It enhances XJoin~\cite{XJOIN} operator and combines it with Symmetric Hash Join~\cite{Deshpande:2007:AQP:1331939.1331940}. Both of them are non blocking operator that save the retrieved result to hash table. Similar to others frameworks, it also has data catalogue that contains list of predicates. Additionally, execution time-out information of SPARQL Endpoint is added in the data catalogue. Therefore, the data catalogue is updated on the fly. Apart from  updating data catalogue, it also updates the execution plan at runtime. The Defender~\cite{montoya2012dadfqafoe,MontoyaEtAl_COLD2012} in ANAPSID has the purpose to split up the query from SPARQL 1.0 format to SPARQL 1.1 format. Not only splitting up the query, Defender also composes related sub query in the same group by exploiting bushy tree. 

\item[SemWIQ] \hfill \\
SemWIQ is another system building on top of ARQ and part of the Grid-enabled Semantic Data Access Middleware (G-SDAM). It provides a specific wrapper to allows data source without equipped SPARQL Endpoint connected. The query federation relies on data summaries in RDFStats and SDV\footnote{\url{http://purl.org/semwiq/mediator/sdv#}}. RDFStats is always up-to-date statistic information since the monitoring component periodically collects information at runtime and stores it into a cache. As the RDFstats also covers histogram String, Blank Node etc, it is more beneficial for SemWIQ to be able to execute any kind of query pattern. 
SDV is based on VoID which is useful for data source registration. The query is parsed by Jena SPARQL processor ARQ before optimization process. SemWIQ applies several query optimisation methods based on statistic cost estimation such as push-down of filter expressions, push down of optional group patterns, push-down of joins and join and union reordering. During optimization, the federator component inserts SERVICE keyword and SPARQL Endpoint for each sub query.

\item[WoDQA] \hfill \\
WoDQA (Web of Data Query Analyzer)~\cite{WoDQA2012} also uses ARQ as a query processor. The source selection is done by analysing metadata in the VoID stores such as CKAN\footnote{http://ckan.net/} and VoIDStore\footnote{http://void.rkbexplorer.com/}. The source observation is based on Internationalized Resource Identifier (IRI), linking predicate and shared variables. It does not exploit any statistic information in the VoID of each dataset, but it only compares IRI or linking predicate to subject, predicate and object. The same variables in the same position are grouped in one sub query. After detecting relevant sources for each subquery, the SERVICE keyword is appended following with SPARQL Endpoint address.

\end{description}



\begin{table}
\caption{The Existing Frameworks Supports Federation over SPARQL Endpoints, Reformulate query to SPARQL 1.1.}
\label{tab:currentdatafed11}
\begin{tabular}{|p{1.5cm}|p{2cm}|p{1.5cm}|p{2.2cm}|p{1cm}|p{2cm}|p{1.6cm}|p{0.6cm}|}
\hline 
Framework & Catalogue & Platform & Source Selection & Cache  & Query Execution & Source Tracking & GUI\\ 
\hline 
SemWIQ & RdfStats+VoID & Jena & Statistic + Service & \tickYes & Bind Join & Dynamic & \tickYes \\ 
\hline 
Anapsid & Predicate List and Endpoint status & Anapsid & Predicate List & \tickNo & Symmetric Hash Join and XJoin &  Dynamic & \tickYes \\ 
\hline 
WoDQA & VoID Stores & Jena & List of predicates and ontologies & \tickNo & \tickNo & Dynamic & \tickYes \\ 
\hline 
\end{tabular} 
\end{table}

\section{Desired Features}
\label{sec:desiredfeatures}
We have seen existing federation SPARQL frameworks along with their behaviours and properties. Based on our summary and experience, we suggest several features that could be added into their framework.
\begin{itemize}
\item Hybrid data catalogue\hfill \\
As described in the section~\ref{fed-exist}, the data source registration could be done by the mediator as well as third party such as search engine. In term of querying in the Linked Open Data, the data source registered should be not limited. The framework could combine static and dynamic data source registration where the data source in the static registration is given higher priority than data source in the dynamic registration before delivering a query. 
\item Link Predicate Awareness \hfill 
 
Link predicate has ability to connect one entity to another entity in a different datasource. For instance, Figure~\ref{fig:dbpediadrugbankkegg} shows the \url{drugbank:keggCompoundID} link joining the entity \url{drugbank:drugs} in the Drugbank dataset with the class \url{kegg_resource:Compound} in the Kegg dataset and \url{owl:sameAS} link joining the entity \url{dbpedia-owl:Drug} in the DBpedia dataset with the entity \url{drugbank:drugs} in Drugbank. Assuming the link predicate connects two datasets, we should avoid to deliver the same destination of query patten containing the link predicate. In the case of Query~\ref{linkpresparql}, pattern \textit{?compound rdfs:label ?compoundname} should not be send to Drugbank dataset as \url{drugbank:keggCompoundID} is a link predicate, even predicate \textit{rdfs:label} occurs in all datasets. 

\lstset{language=SPARQL, caption=Example of Link Predicate Problem, label=linkpresparql,frame=single,breaklines=true,basicstyle=\small}
\begin{lstlisting}[float]
SELECT ?drug ?compoundname 
WHERE {
?drug drugbank:keggCompoundID ?compound .
?compound rdfs:label ?compoundname .
}
\end{lstlisting} 

\end{itemize}

\section{Challenges}
\label{Challenges}
According to our investigation in Section~\ref{fed-exist} we note several challenges to be addressed in the future of federation framework development. Federation over SPARQL Endpoints has become actively developed over the last four years, in particular source selection part. This field area is still infancy which faces several challenges that need to be tackled :

\begin{itemize}
\item Data Access and Security \hfill \\
The data source and mediator are usually located in different locations, therefore the secure communication process among mediator and data source should be concerned. In addition, updating data should be supported in the SPARQL 1.1. Several SPARQL Endpoints provide authentication feature in order to restrict query access for limited user. However, the unauthorized interception  between mediator and data source have not been undertaken by any federation frameworks yet. The public key cryptography could be implemented in the federation frameworks where mediator and data source share public and private key for data encryption during interaction.
\item Data Allocation \hfill \\
Since several RDF stores crawl data from other data source, the data redundancy could not be avoided in the Linked Open Data Cloud. Consequently, the federation over SPARQL Endpoint framework detects data source is located in multiple location. This such condition could increase communication cost during selection source and query execution stage particularly for federation system employing data statistic from third party. Furthermore, the redundancy data could increase intermediate results as more data duplication from multiple source. On the one hand, using popular vocabulary allows user to query easily, but on the other hand, the source prediction for a query will be a hard task. As pointed out by~\cite{6337113} when popular entities and vocabularies are distributed over multiple data sources, the performance of federation query is getting worse.
\item Data Freshness\hfill \\
The freshness is one most important measurement in the data integration because each data source might has different freshness value. Having up to date data catalogue is a must in the federation framework to achieve high freshness value. Inaccurate results could arise from inaccurate data catalogue. Nevertheless, updating data catalogue is a costly operation in term of query execution and traffic between data source and frameworks. Apart from data catalogue being static, the freshness could not be obtained when the high network latency occurs during communication process. 
\item Benchmark\hfill \\
To date, benchmarks are generally proposed for single RDF stores such as LUBM~\cite{Guo2005158}, BSBM~\cite{Bizer09theberlin}, and SP2Bench~\cite{DBLP:journals:corr:bs-0806-4627}. Hence, they are not suitable for distributed infrastructure. FedBench~\cite{DBLP:conf:semweb:SchmidtGHLST11} is the only benchmark proposed for federated query which evaluated the federated query infrastructure performance including loading time and querying time. Those performance metrics are lack to evaluate federation framework. The federation framework benchmark should take into account several performance measurements from traditional distributed database such as query throughput. In addition, several metrics particularly occurring in the federation framework should be considered. For instance, the size of intermediate result, number of request, amount of data sent, etc.   Apart from performance metric, due to heterogeneous data in the federation query, the evaluation of data quality become important measurement namely freshness, consistency, completeness  and accuracy. \cite{montoya2012dadfqafoe} added two more FedBench measurements, namely Endpoint Selection time and completeness. Furthermore, it evaluated performance federation framework in various environment.
Since the FedBench has static dataset and query set, it is difficult task to evaluate framework for other dataset. To address this problem, SPARQL Linked Open Data Query Generator (SPLODGE)~\cite{DBLP:conf:semweb:GorlitzTS12} generates random query set for specified dataset. The query set generation is based on dataset characteristic that is obtained from its predicate statistic. Beside dataset characteristic, it also considers the query structure and complexity such as number of join, the query shape, etc to produce the query set. 
\item Overlapping Terminologies\hfill \\
The data is generated, presented and published using numerous expressions, namespaces, vocabularies and terminologies, that significantly contain duplicate or complementary data~\cite{quackenbush2006standardizing,bechhofer2011linked}. As an example, there are multiple datasets in the LSLOD describing the concept \textit{Molecule}- in Bio2RDF’s kegg dataset, it is represented using \textit{kegg:Compound} whereas in chebi, these are identified as \textit{chebi:Compound} and in BioPax they are denoted as \textit{biopax-level3.owl\#SmallMolecule}, i.e, using different vocabularies and terminologies to explain the similar or related concepts \cite{HasnainOEDW2012} . Moreover different dataset contains different fractions of data or predicates about the same entities e.g: Chebi dataset contains data regarding the mass or the charge of a Molecule whereas Kegg dataset explains Molecule’s interaction with biological entities such as Proteins. Conceptual overlap and different datasets share data about the same entities in LD4LS can be seen in the Figure \ref{fig:LSDomain} . Therefore, the  mapping rules among heterogeneous schemas is highly required in federation query. This task could be done by having global schema catalogue that maps related concepts or properties and generating more links among related entities.
\item Provenance\hfill \\
As more than one datasets involves in Federated SPARQL Query, the origin of data result will be prominence. Apart from number of sources, data redundancy often occurs in the Federation SPARQL query, particularly in Federation over Linked Open Data. It is because several publishers expose the same dataset. For example, Sindice contains DBpedia data: while a user is requesting DBpedia data, the DBpedia and Sindice SPARQL Endpoints are able to answer that query. The redundant result can not be avoided by Federation Framework using third party catalogue. Hence, the data provenance is important factor in the Federation over SPARQL Endpoint. \cite{ekawHarland12} explains a notable provenance implementation in the Federation System called OPENPHACTS\footnote{\url{http://www.openphacts.org/}}. In order to tackle provenance issue, OPENPHACTS utilizes a Nanopublication~\cite{Groth:2010:AN:1883685.1883690} which supports provenance, annotation, attribution and citation.

\begin{figure}
  \centering
	\includegraphics[width=.9\linewidth]{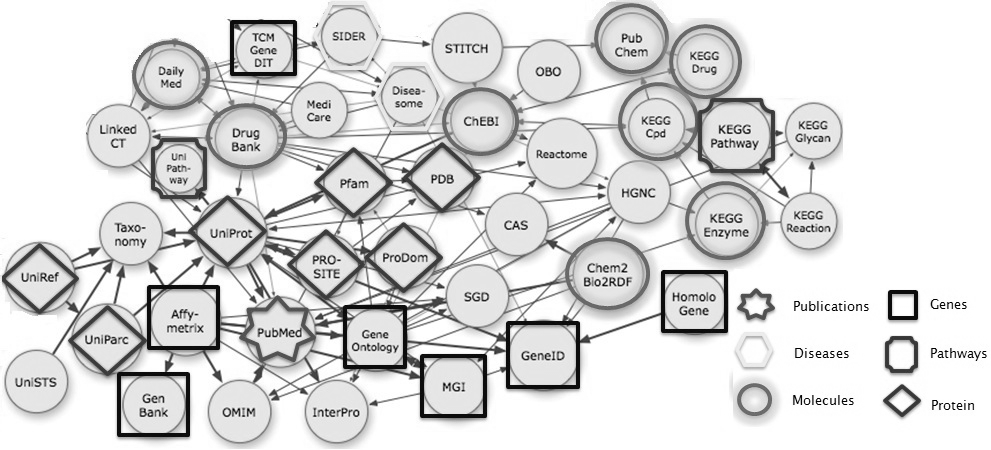}	
	\caption{Different Life Science Datasets talks about same concepts}
	\label{fig:LSDomain}
\end{figure}

\end{itemize}
\section{Conclusion}
\label{conclusion}
Federation query over SPARQL Endpoints made a significant progress in the recent years. Although a number of federation frameworks have already been developed, the field is still relatively far from maturity. Based on our experience with the existing federation frameworks, the frameworks mostly focus on source selection and join optimization during query execution. 

In this work, we have presented a list of federation frameworks over SPARQL Endpoints along with their features. According to this list, the user can have considerations to choose the suitable federation framework for their case. We have classified those framework into three categories: i) framework interprets SPARQL 1.1 query to execute federation SPARQL query covering \textit{VALUES} and \textit{SERVICE} operator; ii) framework handles SPARQL 1.0 query and has responsibility to find relevant source for a query and join incoming result from SPARQL Endpoints; and iii) framework accepts SPARQL 1.0 and translate the incoming query to SPARQL 1.1 format. 

Based on the current generation of federation frameworks surveyed in this paper, it still requires further improvements to make frameworks more effective in a broader range of applications. We suggested several features that could be included in the future developments. Finally, we point out challenges for future research directions.

\bibliographystyle{alpha}
\bibliography{kesw-extended}

%
%

\end{document}